\documentclass[preprint,12pt]{elsarticle}
\journal{Computer Physics Communications}

\usepackage[T1]{fontenc} 
\usepackage{framed}

\usepackage[utf8]{inputenc}
\usepackage{url}
\usepackage{graphicx}
\usepackage{axodraw2}

\newcommand{\be}{\begin{equation}}
\newcommand{\ee}{\end{equation}}
\newcommand{\bea}{\begin{eqnarray}}
\newcommand{\eea}{\end{eqnarray}}
\newcommand{\dd}{\mbox{d}}
\newcommand{\pa}{\partial}
\newcommand{\al}{\alpha}

\newenvironment{code}
  {\tt\noindent\begin{framed}\begin{flushleft}}
  {\end{flushleft}\end{framed}\normalfont}

\begin{document}

\begin{frontmatter}

\title{FIRE6: Feynman Integral REduction with Modular Arithmetic}

\author[SRCC,KIT]{A.V.~Smirnov\corref{cor1}}
\ead{asmirnov80@gmail.com}

\author[CS]{F.S.~Chukharev}
\ead{fedor.s.chukharev@gmail.com}

\cortext[cor1]{Corresponding author}

\address[SRCC]{Research Computing Center, Moscow State University, \\ 119992 Moscow, Russia}
\address[KIT]{Institut f\"{u}r Theoretische Teilchenphysik, Karlsruhe Institute of Technology (KIT), \\ 76128 Karlsruhe, Germany}
\address[CS]{Faculty of Computational Mathematics and Cybernetics, Moscow State University, \\ 119992 Moscow, Russia}

\begin{abstract}
{\tt FIRE} is a program performing reduction of Feynman integrals to master integrals.
The {\tt C++} version of {\tt FIRE} was presented in 2014.
There have been multiple changes and upgrades since then including
the possibility to use multiple computers for one reduction task and to perform reduction with modular arithmetic.
The goal of this paper is to present the current version of {\tt FIRE}.
\end{abstract}
\end{frontmatter}
\newpage

{\bf PROGRAM SUMMARY}

\vspace{1cm}

\begin{small}
\noindent
{\em Manuscript Title:} FIRE6 Feynman Integral REduction with Modular Arithmetic\\
{\em Authors:} A.V. Smirnov, F.S. Chukharev\\
{\em Program title:} FIRE, version 6 (FIRE6)\\
{\em Licensing provisions:} GPLv2\\
{\em Programming language:} {\tt Wolfram Mathematica} 6.0 or higher, {\tt C++}\\
{\em Computer(s) for which the program has been designed:} starting from a desktop PC up to a supercomputer\\
{\em Operating system(s) for which the program has been designed:} Linux 64bit, Mac OS X 10.6 or higher 64bit\\
{\em RAM required to execute with typical data:} depends on the complexity of the problem  \\
{\em Has the code been vectorized or parallelized?:} yes\\
{\em Number of processors used: } depending on the mode and the complexity of the task, the program can both run single-threaded on a laptop, or use a supercomputer (tested with up to 2024 cores) \\
{\em Supplementary material:} The article, install instructions, \\https://bitbucket.org/feynmanIntegrals/fire
{\em CPC Library Classification:} 4.4 Feynman diagrams, 4.8 Linear Equations and Matrices, 5 Computer Algebra, 20 Programming and Publication Practice \\
{\em External routines/libraries used:} {\tt Wolfram Mathematica} [1], {\tt Snappy} [2], {\tt ZStandard} [3], {\tt KyotoCa\-binet} [4], {\tt Fermat} [5], {\tt LiteRed} [6] \\
{\em Nature of problem:}
Reducing Feynman integrals to master integrals can be treated as a task to solve
a huge system of sparse linear equations with polynomial coefficients.\\
{\em Solution method:}
Since the matrix of equations is very specific, none of standard methods of solving linear 
equations can be applied efficiently. The program approaches solving those equations 
with a special version of Gauss elimination.
In complex cases the direct reduction approach might fail, so the approach with modular arithmetic is used,
where the reduction is performed multiple times with different values of variables over large prime number fields,
afterwards the coefficients are reconstructed.
The data preparation and result analysis is performed in {\tt Wolfram Mathematica} [1],
but the main reduction procedure is written in {\tt C++};
{\tt FIRE} compresses data with the use of the {\tt Snappy} [2] or {\tt ZStandard} [3] library,
stores it on disk with the use of the {\tt KyotoCabinet} [4] database engine,
and performs algebraic simplifications with the {\tt Fermat} [5] program.
The external package {\tt LiteRed} [6] can be used to produce additional
rules for reduction. \\
{\em Restrictions:} The complexity of the problem is mostly restricted
by CPU time required to perform the reduction of integrals and the available RAM.
The program has the following limits: maximal number of indices = 22, maximal number of positive indices = 15, 
maximal number of non-trivial sectors = $128\times256-3=32765$ (global symmetries decrease the number of sectors, indices that cannot be positive
do not double the number of sectors).
{\tt FIRE6} follows the {\tt C++11} standard, so requires {\tt gcc 4.8.1} or higher to be compiled, but works with the current {\tt gcc 7.3} as well.
\\
{\em Running time:} depends on the complexity of the problem\\
{\em References:} 
{\\}
[1] \url{http://www.wolfram.com/mathematica/}, commercial algebraic software; 
{\\} [2] \url{https://github.com/google/snappy}, open source;
{\\} [3] \url{https://github.com/facebook/zstd}, open source;
{\\} [4] \url{http://fallabs.com/kyotocabinet/}, open source;
{\\} [5] \url{https://home.bway.net/lewis/}, free--ware with some restrictions for organizations;
{\\} [6] \url{http://www.inp.nsk.su/~lee/programs/LiteRed/}, open source.

\end{small}

\newpage

\section{Introduction}

Feynman integrals are fundamental objects arising in modern elementary particle physics.
The problems being considered currently result in
millions of Feynman integrals that have to be evaluated.
A classical approach is to apply
\textit{integration by parts (IBP) relations}~\cite{Chetyrkin:1981qh}  
(see Chapter~6 of \cite{Smirnov:2013ym} for a review)
and reduce all integrals to a smaller set,
the so-called \textit{master integrals}\footnote{It has been shown in \cite{Smirnov:2010hn} that the number of master integrals is
always finite.}. 
The first approach to integration by parts relations was based on solving those equations by hand.
This procedure was used in many papers starting from~\cite{Chetyrkin:1981qh}
before the appearance of computer codes to solve IBP relations.

Currently there is a number of programs that can perform Feynman integral reduction,
some of those are public, some are private.
Most of the existing programs are based on the Laporta algorithm~\cite{Laporta:2001dd} (Gauss elimination after choosing an ordering),
in particular, the public codes described in~\cite{Anastasiou:2004vj,Studerus:2009ye,vonManteuffel:2012np,vonManteuffel:2012je,Maierhoefer:2017hyi,Maierhofer:2018gpa}.
However there were also attempts based on constructing explicit rules when solving IBP relations, both
by one of the authors of this paper~\cite{Smirnov:2006tz,Smirnov:2006wh,Smirnov:2007iw}, as well as by R.Lee in his program {\tt LiteRed}~\cite{Lee:2008tj,Lee:2012cn,Lee:2013mka}.

A recent approach to the reduction is based on modular arithmetic. While the direct reduction might be too complex because of the growth of coefficients,
one can fix different values of variables and also perform reduction over fields of remainders over large prime numbers instead of rationals.
Those numbers fit into machine-size integers, so the reduction can run significantly faster. However, to recover coefficients afterwards one has to
run a rather large number of reduction jobs. This approach has already been presented in reduction programs Finred~\cite{vonManteuffel:2016xki} and Kira~\cite{Maierhoefer:2017hyi}.

One of the authors of this paper created the program {\tt FIRE} (Feynman integral reduction) about ten years ago.
The initial version of {\tt FIRE}~\cite{Smirnov:2008iw,Smirnov:2013dia} was written in {\tt Wolfram Mathematica},
later a {\tt C++} version was also published~\cite{Smirnov:2014hma}.

The goal of this paper is to present the new version of {\tt FIRE} that was developed during recent years.
We consider {\tt FIRE6} to be a major release significantly differing from the previous public version.
It was already successfully applied in~\cite{Henn:2016men,Lee:2016ixa,Lee:2017mip,Kurz:2015bia,Marquard:2016dcn,Marquard:2018rwx} ---
the result obtained in those papers were most probably unachievable with the previous versions of FIRE.
In particular, the first example of a calculation which was made real by the use of FIRE with modular arithmetic can be found in \cite{Lee:2019zop}.

The new version brings a performance upgrade, being about twice as fast than the previous public version on small tests.
However the main power comes at problems being ``at the edge of science'' since the memory economy and stability features
results in a ``can be done --- cannot be done'' difference.

Along with a performance upgrade, the new version contains a number of new features, including but not limited to the following:

\begin{itemize}
 \item usage of internal sector symmetries from {\tt LiteRed};
 \item possibility to recover from system crashes without a restart from the very beginning;
 \item possibility to use multiples nodes on one reduction task;
 \item modular arithmetic approach;
 \item MPI-approach to run multiple modular arithmetic tasks on a supercomputer;
\end{itemize}

In section 2 we recall basic definitions related to Feynman integrals, in section 3 we describe how to install {\tt FIRE}
and in section 4 how to use it. Section 5 explains the internals of {\tt FIRE} which might be important for choosing options
efficiently, section 6 is devoted to the usage of modular arithmetic, and in the appendix we summarizes all options of config files.

This paper can also serve as a user manual for {\tt FIRE6}. Additional information and examples can be found in the {\tt FIRE} distribution.
The new version of {\tt FIRE} (starting with {\tt 6.1}) is also covered with the doxygen documentation which can be usefull
for contributors, bug finding or those who wish to use some internal functions. 
Details how to build and read this documentation are also provided below.

\section{Basic definitions}

Let us remind the basic definitions that will be used in this paper.
Feynman integrals are functions of integer variables which are also called indices,
\bea
  F(a_1,\ldots,a_n) &=&
  \int \cdots \int \frac{\dd^d k_1\ldots \dd^d k_h}
  {E_1^{a_1}\ldots E_n^{a_n}}\,.
  \label{eqbn-intr}
\eea
Here the factors $E_i$ in the denominator are linear functions with respect to
scalar products of loop momenta $k_i$ and external momenta $p_i$; dimensional regularization with
$d=4-2\epsilon$ is applied.

One considers the so-called integration by parts relations~\cite{Chetyrkin:1981qh} 
\bea \int\ldots\int \dd^d k_1 \dd^d k_2\ldots
\frac{\pa}{\pa k_i}\left( p_j \frac{1}{E_1^{a_1}\ldots E_n^{a_n}}
\right)   =0   \label{RR-intr}
\eea
They can be rewritten as:
\begin{equation}
\sum \al_i F(a_1+b_{i,1},\ldots,a_n+b_{i,n})
=0\,.
\label{IBP-intr}
\end{equation}
where $b_{i,j}\in \{-1, 0, 1\}$ and $\al_i$ are linear functions of $a_j$.

It is already standard to use the notion of sectors during reduction.
There are $2^n$ sectors, each of those defined by specifying a subset of indices that have to be positive
(and the remaining have to be non-positive).
Each sector has a unique so-called \textit{corner integral}, that is the one with indices equal to $0$ or $1$.

We will define a sector to be  \textit{lower} than another sector if all indices of the corner integral in the first one are smaller than the corresponding
indices of the corner integral in the second one. Basically integrals in lower sectors are simpler, 
so starting from the time of reduction ``by hand'' one tries to reduce Feynman integrals to ones in lower sectors.

We will call a sector \textit{trivial} if all integrals corresponding to sets of indices in this sector are equal to zero. 
It is known that the lowest sector, where all indices are non-positive, is trivial, but in fact a large number of sectors are trivial.
The conditions determining whether a sector is trivial are called \textit{boundary conditions}.

\section{Installation}

{\tt FIRE6} is distributed via bitbucket. 
One can also download a binary package compiled on Ubuntu 16.04 or openSUSE 15.0 from the download section of the repository 
(\url{https://bitbucket.org/feynmanIntegrals/fire/}), but the recommended way is to build {\tt FIRE} from sources.

To do it one has to

1) clone it with {\tt git} (this should be done in a folder that has no space symbols in its full path)

\begin{code} 
git clone https://bitbucket.org/feynmanIntegrals/fire.git
\end{code}

As a result a {\tt fire/FIRE6} folder will appear. The {\tt fire} folder name can be later changed if needed,
however it is not recommended to change the name of the internal folder {\tt FIRE6} to be able to receive updates with {\tt git pull}.
So now one can move to the internal folder with {\tt cd fire/FIRE6}.

To receive updated one can pull them from the repository

\begin{code} 
git pull
\end{code}

For details please refer to git manuals since this paper cannot cover all those.

2) configure {\tt FIRE}. There is a {\tt configure} file in the folder that can be run with 

\begin{code} 
./configure 
\end{code}

This file is not a part of the autoconf system, but a simple script that sets some preprocessor variables and makefiles corresponding to the settings provided to {\tt ./configure}
There are the following options:
\begin{itemize}
 \item {\tt $--$enable\_zlib}: enables the zlib compressor shipped with kyotocabinet, requires the zlib/deflate library (zlib1g-dev) to be installed system-wide;
 \item {\tt $--$enable\_snappy}: tries to build the snappy library shipped with {\tt FIRE} to serve as a compressor;
 \item {\tt $--$enable\_zstd}: tries to build the zstandard library shipped with {\tt FIRE} to serve as a compressor;
 \item {\tt $--$enable\_debug}: adds -g to compile options for debugging symbols and also attaches hooks for printing out stack trace in case of crashes;
 \item {\tt $--$enable\_tcmalloc}: compiles google perf tools in order to make {\tt FIRE} use tcmalloc instead of the libc malloc;
 \item {\tt $--$enable\_lthreads}: gives {\tt FIRE} possibility to use multiple threads inside a sector".
\end{itemize}

The first two options are related to compressor settings of the database engine. By default {\tt FIRE6} uses the {\tt lz4} compressor which comes shipped with {\tt FIRE} and is easiest to build. The {\tt $--$enable\_debug} option is self-explanatory, the {\tt $--$enable\_tcmalloc} option might provide significant memory economy and the {\tt lthreads} option provides a better parallelization resource (details will be explained below).

The options of {\tt configure} are saved in the {\tt previous\_options} file.

3) build the libraries that are shipped with {\tt FIRE}

\begin{code} 
make dep 
\end{code}

This command builds {\tt kyotocabinet}, {\tt lz4}, {\tt snappy} (in the case where {\tt $--$enable\_snappy} is set),
{\tt zstandard} (in the case where {\tt $--$enable\_zstd} is set) and {\tt google perf tools} (if configured with the {\tt $--$enable\_tcmalloc} option). As usual, one can use the {\tt -j N} option where {\tt N} is the number of cores to run compilation faster. As a result, the libraries are ``installed'' in the {\tt usr} subfolder of {\tt fire/FIRE6}.

4) build {\tt FIRE}

\begin{code} 
make 
\end{code}

the simlinks to binaries are located in the {\tt bin} folder. {\tt FIRE} builds a number of binaries, with or without {\tt p} at the end of the name. The {\tt p} binaries are the modular arithmetic version of {\tt FIRE} (comes from ``prime''). The objects and binaries themselves are created in the {\tt poly} and {\tt prime} folders. Both of them contain the {\tt FIRE6}, {\tt FLAME6} and {\tt FTool6} binaries. The {\tt FIRE6} is the main binary, {\tt FLAME6} is used for sector jobs and {\tt FTool6} is an auxiliary binary that can be used for database analysis.

5) run tests

\begin{code}  
make test 
\end{code}

The new version of {\tt FIRE} has a set of tests that are also available at bitbucket via bitbucket pipelines. 

6) To build the {\tt mpi} version of {\tt FIRE} (for use on supercomputers) one has to run

\begin{code} 
make mpi 
\end{code}

This version can also be tested on a personal computer in case the {\tt openmpi} libraries are installed system-wide. The {\tt mpicxx} binary is called for this compilation,
this can be changed by editing {\tt mpi/Makefile}.

7) The development of {\tt FIRE} now follows the gitflow strategy. The master branch only receives hotfixes and is moved mostly when a new release is being prepared. 
Both the master and the dev branches are changed only with pull requests, and a bunch of tests is performed via bitbucket pipelines before such a pull request can be merged.
The dev branch however can receive changes much faster, so if one is willing to receive updates between releases, the dev branch has to be chosed.

\begin{code} 
git checkout dev
\end{code}

After that everything should be recompiled.

8) The {\tt FLink} and {\tt KLink} binaries are no longer a part of the {\tt FIRE} distribution, so there are no more instructions on how to build them.

9) The doxygen documentation can be built with

\begin{code} 
make doc

make pdf
\end{code}

The first command builds doxygen html and latex documentation. It requires doxygen to be installed. 
The mein generated index file of the html documenntation is {\tt \url{documentation/html/index.html}}. 
The second command can also build a pdf manual from the latex version and required pdflatex to be installed. 
If successful, it created the {\tt documentation/refman.pdf} file.
If the {\tt xdg-open} utility is installed, the documentation in its html format can be also opened with

\begin{code} 
make showdoc
\end{code}

\section{Basic usage of {\tt FIRE}}

{\tt FIRE} has a part written in {\tt Wolfram Mathematica} and a much bigger part written in {\tt C++}. While {\tt FIRE} is able to perform reduction by the {\tt Mathematica} part, it should be used mainly for tests and for simple tasks. Partially because of that the support of calls to {\tt fermat} and the database engine from {\tt Mathematica} were removed in the current version. If a problem is complex enough that {\tt Mathematica} cannot handle it with ease, one should use the {\tt C++} part. 

The proper approach to use {\tt FIRE} for complicated problems is the following. First one should create a start file in {\tt Mathematica}. Then this file is used by the {\tt C++} part to run the reduction, afterwards tables are created which can be read by {\tt Mathematica} in order to produce results. Moreover it is strictly recommended to use {\tt LiteRed} in order to produce symmetries before running the reduction.

\subsection{Preparing a start file}

The {\tt Mathematica} part of {\tt FIRE} is loaded simply with

\begin{code} 
Get["FIRE6.m"]; 
\end{code}
 
\noindent provided that the currect directory is set to {\tt fire/FIRE6}. The recommendation to set the {\tt FIREPath} variable used in previous versions is no longer valid. 

Like in previous versions one has to prepare a start file to use {\tt FIRE}. Such a file contains information on the reduction of one family of Feynman integrals --- dimension, sectors, integration-by-parts relations.
To have backward compatibility we keep the syntax at this stage. One has to set the following variables.

\begin{itemize}
 \item {\tt Internal} --- the list of internal momenta, for example, \{k\};
 \item {\tt External} --- the list of external momenta, for example, \{p1, p2, p4\}; one should list only linearly independent external momenta after having used momentum conservation;
 \item {\tt Propagators} --- the list of propagators, for example, \{-k$^2$, -(k + p1)$^2$, -(k + p1 + p2)$^2$, -(k + p1 + p2 + 
       p4)$^2$\}; the propagators should be quadratic in terms of momenta;
 \item {\tt Replacements} (optional) --- the list of replacement rules for kinematic invariants, for example, \{p1$^2$ -> 0, p2$^2$ -> 0, p4$^2$ -> 0, p1 p2 -> s/2, 
   p2 p4 -> t/2\};
\end{itemize}

There are also two more variables that were usually set in previous versions but can be skipped now.

\begin{itemize}
 \item {\tt RESTRICTIONS} (optional) --- list of boundary conditions. For example if this list has an element {\tt \{-1, -1, -1, 0\}}, this means that the integrals are equal to zero if the first three indices are non-positive. Since currently {\tt FIRE} can detect boundary conditions automatically in most cases, this option can be skipped. Moreover, the usage of {\tt LiteRed} can detect even more boundary conditions at a later stage. Still, if this variable is set, those restrictions are also used;
 \item {\tt SYMMETRIES} (optional) --- list of symmetries (permutations of indices not changing the integrals, each element consists of list positions); in older versions one had to provide the whole symmetry group, but currently it is enough to provide the generators; for example, if this list has an element {\tt \{3, 2, 1, 4\}}, this means that {\tt F[a,b,c,d]=F[c,b,a,d]}. In principle, the use of symmetries can speed up the reduction, however in case {\tt LiteRed} is used, it can also detect equivalent sectors and create corresponding mapping rules. This will be somewhat slower compared to the use of global symmetries, but we estimate the difference as relatively small.
\end{itemize}

If one wishes to set the integration by parts relations manually, there is a special variable {\tt startinglist} that should be set (see \cite{Smirnov:2014hma} for details).

Now one has to run the {\tt PrepareIBP[]} and then {\tt Prepare[]} commands. The second one leads also to the autodetection of boundary conditions. This can be turned off by running 

\begin{code} 
Prepare[AutoDetectRestrictions -> False] 
\end{code}

\noindent instead. The command also has a parallel option run with 

\begin{code} 
Prepare[Parallel -> True]
\end{code}

Now a start file can be saved with {\tt SaveStart["filename"]}. It is recommended to quit the kernel afterwards.

\begin{figure}[ht!]
\centering
\includegraphics[width=90mm]{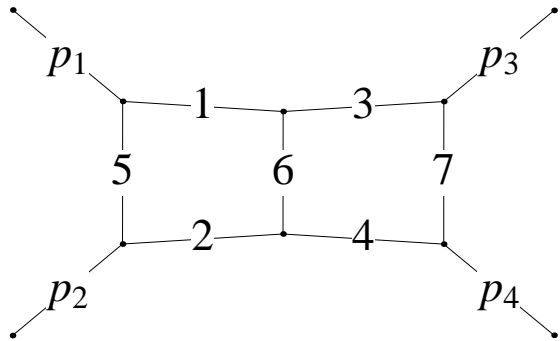}
\label{doublebox}
\caption[]{Massless on-shell doublebox}
\end{figure}

For example, for a double box diagram with massless lines, incoming momenta {\tt p1}, {\tt p2} and {\tt p3} and {\tt p1$^2$ = 0, p2$^2$ = 0, p3$^2$ = 0, (p1+p2+p3)$^2$ = 0, s = (p1 + p2)$^2$, t = (p1 + p3)$^2$} one has:

\begin{code}
Get["FIRE6.m"];
     
Internal = \{k1, k2\};

External = \{p1, p2, p3\};

Propagators = \{-k1$^2$, -(k1 + p1 + p2)$^2$, -k2$^2$, -(k2 + p1 + p2)$^2$, -(k1 + p1)$^2$, -(k1 - k2)$^2$, -(k2 - p3)$^2$, -(k2 + p1)$^2$, -(k1 - p3)$^2$\};
       
Replacements = \{p1$^2$ -> 0, p2$^2$ -> 0, p3$^2$ -> 0, p1 p2 -> s/2, 
   p1 p3 -> t/2, p2 p3 -> -1/2 (s + t)\};       
       
PrepareIBP[];

Prepare[];

SaveStart["doublebox"];

Quit[];
\end{code}

As a result one has a file {\tt "doublebox.start"} containing all the required information to proceed.

Please note that start files created by old versions of {\tt FIRE} (public versions up to 5.2 and some private versions) cannot be used by the current {\tt C++} part. 
They have to be either recreated or converted by the {\tt ConvertStart[oldfile, newfile]} command.

\subsection{Reduction in {\tt Wolfram Mathematica}}

The start file can be loaded in {\tt Mathematica} with the {\tt LoadStart} command. For example, 

\begin{code}
Get["FIRE6.m"];

LoadStart["doublebox", 1];

Burn[]
\end{code}

The second argument is the number assigned to the current family of Feynman integrals. It is recommended to have different numbers for different families, this will
allow one to distinguish integrals belonging to different families and to find equivalents between them later. This should be a positive number fitting an unsigned short integer (less than $2^{16}$).

In complicated situations {\tt Burn[]} might work slowly. Hence one can run
{\tt SaveData["filename"} after {\tt Burn[]}, then quit the kernel and later
load everything with {\tt LoadData["filename"]} (without {\tt LoadStart} and {\tt Burn}).

Now one can perform the reduction with

\begin{code}
F[1, \{1, 1, 1, 1, 1, 1, 1, -1, -1\}]
\end{code}

Here {\tt F} is a call to perform the reduction, ``{\tt 1}'' is the same number as in {\tt LoadStart} and {\tt \{1, 1, 1, 1, 1, 1, 1, -1, -1\}} is the set of indices.
As a result one has an expression of the form

\begin{code}
3/2 s G[1,\{1,1,1,1,1,1,1,-1,0\}] + 1/2 s t G[1,\{1,1,1,1,1,1,1,0,0\}] + \ldots
\end{code}

\noindent where the dots refer to further terms that have been omitted for brevity.
In the resulting expression {\tt G} follows the same notation as {\tt F} but is treated as an irreducible (master) integral and does not lead to new reduction.

As it has been mentioned earlier, running complicated reductions in {\tt Mathematica} is not recommended. 
But in case this reduction is performed it is strictly recommended to avoid the {\tt F} command which reduced integrals one by one and to use
the more similar to the {\tt C++} mode command {\tt EvaluateAndSave}.
It has two arguments: the list of integrals that has
to be reduced and the file where to save tables. For example, to reduce two integrals by this command one should run

\begin{code} 
EvaluateAndSave[\{\{1, \{1, 1, 1, 1, 1, 1, 1, -1, -1\}\},\\ \{1, \{1, 1, 1, 1, 1, 1, 1, 0, -2\}\}\},"doublebox.tables"]
\end{code}

This command creates the tables file which can be loaded, and then the {\tt F} command can be used to produce results.

\subsection{Loading the tables}

The tables created with the {\tt EvaluateAndSave} or with the {\tt C++} version can be loaded with the {\tt LoadTables} command.
In case multiple tables are to be loaded one should pass all file names in a list to a single call to {\tt LoadTables}.

\begin{code}
LoadTables[\{file1,file2,$\ldots$,fileN\}]
\end{code}

Now a call to {\tt F} will take the result from tables.
Sometimes this can be slow because {\tt FIRE} tries to factorize coefficients in order to present a 
nice-looking result. This behavior can be switched off by {\tt FactorCoefficients = False}.

In case one prefers to use the tables without loading start file and the {\tt F} syntax, they can be converted into {\tt Mathematica} rules.
The command for this is

\begin{code}
Tables2Rules[filename, Func: Identity, JoinTerms: True] 
\end{code}

It is not required to load start files to use this command and get a list of {\tt Mathematica} rules as a result. The first argument is the file name with tables.
The second (optional) argument is the function that is applied to all coefficients, where default value is {\tt Identity} meaning no function
and a reasonable choice can be {\tt Factor} or {\tt Together}.
The third (optional) argument is the indication whether the right-hand side is transformed into a sum of integrals with coefficients.
Setting it to {\tt False} provides a list of pairs (an integral and a coefficient) which can be usefull in case of 
further automatic processing of results.

\subsection{{\tt C++} reduction}

As it has been mentioned a few times above, the reduction should be run with the {\tt C++} part of the program.
As a result it created tables that can be loaded into {\tt Mathematica} afterwards. 

To run the {\tt C++} version, one has to create a configuration file (with the {\tt config} extension).
All options of config files will be discussed below, but let us first illustrate the syntax with this example:

\begin{code}
$\#$variables         d, s, t

$\#$start

$\#$folder            examples/

$\#$problem           1 doublebox.start

$\#$integrals         doublebox.m

$\#$output            doublebox.tables
\end{code}

The spaces are insignificant, each line should start from $\#$ (nowadays called the hash symbol).
$\#${\tt variables} should list all variables that can appear. If variables are set incorrectly, the reduction will freeze due to fermat. 

The $\#${\tt folder} command is optional and provides a path to a folder where files listed by the following commands reside 
(unless specfied with an absolute path). If the folder instruction is missing, the paths are considered absolute or from the ``current'' directory.
$\#${\tt problem} followed by the number corresponding to the current family of Feynman integrals and the path to a start file. 
$\#${\tt integrals} points to a file containing a list of integrals that have to be reduced and $\#${\tt output} points to a file where
the resulting tables are to be saved. The input is a {\tt Mathematica} list of pairs, for example,
{\tt \{\{1, \{1, 1, 1, 1, 1, 1, 1, -1, -1\}\}, \{1, \{1, 1, 1, 1, 1, 1, 1, 0, -2\}\}\}}.

Now the reduction can be launched with 

{\tt bin/FIRE6 -c examples/doublebox}

It should be stated once more that this reduction does not need {\tt Ma\-the\-ma\-ti\-ca}. As a result one gets a file containing 
reduction results --- expressions of integrals via master integrals.

\subsection{Using {\tt LiteRed} rules in {\tt Mathematica}}

{\tt LiteRed} is a program by R. Lee \cite{Lee:2008tj,Lee:2012cn,Lee:2013mka} that aims at solving the IBP relations
before the substitution of indices. This is an alternative approach to the IBP reduction.
However even without reduction rules the {\tt LiteRed} program can provide important information that should be used in {\tt FIRE}.
This information consists of

\begin{itemize}
 \item Boundary conditions (the knowledge all integrals in some sectors are equal to zero);
 \item External symmetries (mappings between different sectors);
 \item Internal symmetries (mappings inside sectors);
 \item Reduction rules in some sectors (we normally avoid this, but sometimes it can help).
\end{itemize}

Currently {\tt FIRE} comes shipped with the version {\tt 1.8} of {\tt LiteRed}. 
It is recommended not to use {\tt LiteRed} downloaded elsewhere since there can be a difference in file formats.

If one wishes to upgrade to the most recent
version of {\tt LiteRed} it can be downloaded from \url{http://www.inp.nsk.su/~lee/programs/LiteRed/},
however we cannot guarantee their compatibility.

Please note that {\tt LiteRed} is not a part of {\tt FIRE} but a program having another author. 
Therefore, when using {\tt LiteRed} options one should give proper credit to {\tt LiteRed} program by citing \cite{Lee:2008tj,Lee:2012cn,Lee:2013mka}.

To use {\tt LiteRed} rules one has to construct them first. 
There are multiple examples shipped with the {\tt LiteRed} package, let us consider one of them, 
the two-loop massless onshell vertex.

\begin{figure}[ht!]
\centering\includegraphics[width=5cm]{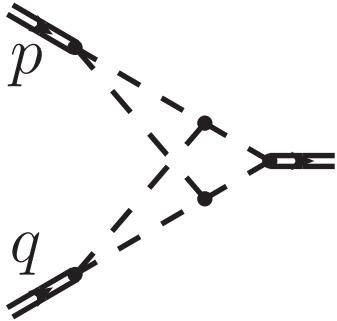}
\caption{Two-loop massless onshell vertex.}
\label{fig:1}
\end{figure}

In the context of {\tt FIRE} it is convenient to call {\tt LiteRed} rules construction with

\begin{code}
 
SetDirectory["extra/LiteRed/Setup/"];

Get["LiteRed.m"];

SetDirectory["../../../"];

Get["FIRE6.m"];

Internal = \{l, r\};

External = \{p, q\};

Propagators = \{-(l - r)$^2$, -l$^2$, -r$^2$, -(-l + p)$^2$, -(q - r)$^2$, \\ -(-l + p + 
    r)$^2$, -(l + q - r)$^2$\};  

Replacements = \{p$^2$ -> 0, q$^2$ -> 0; p q -> -1/2\};
    
CreateNewBasis[v2, Directory -> "temp/v2.dir"];
  
GenerateIBP[v2];

AnalyzeSectors[v2, \{0, \_\_\}];

FindSymmetries[v2,EMs->True];

DiskSave[v2];

Quit[];

\end{code}

Note: {\tt CreateNewBasis} is a not a command of {\tt LiteRed}, it is a command in {\tt FIRE} that translates {\tt FIRE} input into {\tt LiteRed} input.

Here  the {\tt extra/LiteRed/Setup/} is inside the {\tt FIRE6} folder, {\tt v2} is a name fo the diagram used inside {\tt LiteRed}.

The {\tt AnalyzeSectors} command finds trivial sectors. The last parameter in this example assumes the one is considering integrals with non-positive first index.
In other cases one should move the position of the zeros and underscore wildcards.

The {\tt FindSymmetries} command finds symmetric sectors meaning those that can be mapped as sums to other sectors, the {\tt EMs->True} option stands for the autodetection of external symmetries. 
For other syntax variants please refer to {\tt LiteRed} examples.

Before {\tt DiskSave} one can also invoke the {\tt SolvejSector} command in order to construct rules for some sectors.
However our experience shows that one is advised to be careful with it because partial rules can degrade performance of the Laporta approach.

As a result one gets a folder with multiple files containing information on the chosen family of Feynman integrals. They can be loaded in {\tt Mathematica} with

\begin{code}

Get["FIRE6.m"];

LoadStart["examples/v2", 2];

LoadLRules["temp/v2.dir", 2];

Burn[];

\end{code}

Here ``$2$'' stands for the problem number that should be the same in both lines. The {\tt LoadLRules} command reads the directory with {\tt LiteRed} rules and loads 
everything it can use out of there. Then the reduction can be performed as before.

\subsection{Using {\tt LiteRed} rules in {\tt C++}}

The {\tt C++} version is not capable of parsing the {\tt LiteRed} folder directly, they have to be converted first.
It can be done with

\begin{code}
 
Get["FIRE6.m"];

LoadStart["examples/v2"]; 

TransformRules["temp/v2.dir", "examples/v2.lbases", 2];

SaveSBases["examples/v2"];

\end{code}

Here {\tt 2} is the problem number. It is important to mention that the result of the {\tt TransformRules} command is both a {\tt lbases} file and 
a new file with the {\tt sbases} extension that should be used instead of the original start file.
The syntax of those files is currently the same (sbases comes from a deprecated idea to use Groebner bases in
order to run the reduction), however we keep different extensions in order to distingush the original start file and the result of {\tt TransformRules}.
The new file contains some extra information compared with the original start file such as sector priorities, orderings, extra boundary conditions.

Now the config file can be changed and should contain the following lines:

\begin{code}

$\#$problem           2 v2.sbases

$\#$lbases            v2.lbases

\end{code}

It is worth to mention that lbases files created by old versions of {\tt FIRE} (up to public 5.2 and some private versions) do not contain internal symmetries and should be re-created.

\subsection{Finding equivalents between master integrals}

\label{rules}
Extra master integrals that are equivalent to each other can degrade performance a lot for final reductions.
Normally the usage of {\tt Litered} should result in having no equivalent masters. However sometimes they can still exist.
In order to find equivalents one should use the {\tt FindRules} and {\tt WriteRules} commands.

In the doublebox example the {\tt MasterIntegrals[]} command can be used to obtain the list of masters:

\begin{code} 
\{\{1, \{0, 0, 0, 0, 1, 1, 1, 0, 0\}\}, \{1, \{0, 0, 1, 1, 1, 1, 0, 0, 
   0\}\}, \{1, \{0, 0, 1, 1, 1, 1, 1, 0, 0\}\}, \{1, \{0, 1, 1, 0, 0, 1, 0, 0,
    0\}\}, \{1, \{0, 1, 1, 0, 1, 1, 1, 0, 0\}\}, \{1, \{1, 0, 0, 1, 0, 1, 0, 
   0, 0\}\}, \{1, \{1, 0, 0, 1, 1, 1, 1, 0, 0\}\}, \{1, \{1, 1, 0, 0, 0, 1, 1,
    0, 0\}\}, \{1, \{1, 1, 0, 0, 1, 1, 1, 0, 0\}\}, \{1, \{1, 1, 1, 1, 0, 0, 
   0, 0, 0\}\}, \{1, \{1, 1, 1, 1, 1, 1, 1, 0, 0\}\}, \{1, \{1, 1, 1, 1, 1, 1,
    1, -1, 0\}\}\}
\end{code} 

It is easy to see that some of the integrals in lower sectors are equivalent.
This can be found with 

\begin{code} 

Internal = \{k1, k2\};

External = \{p1, p2, p3\};

Propagators = \{-k1$^2$, -(k1 + p1 + p2)$^2$, -k2$^2$, -(k2 + p1 + 
       p2)$^2$, -(k1 + p1)$^2$, -(k1 - k2)$^2$, -(k2 - p3)$^2$, -(k2 + 
       p1)$^2$, -(k1 - p3)$^2$\};
       
Replacements = \{p1$^2$ -> 0, p2$^2$ -> 0, p3$^2$ -> 0, p1 p2 -> s/2, \\
   p1 p3 -> t/2, p2 p3 -> -1/2 (s + t)\};       

FindRules[MasterIntegrals[]]   
   
\end{code} 

or saved to a file with

\begin{code} 
WriteRules[MasterIntegrals[],  "examples/doublebox"];
\end{code} 

The resulting file has a syntax like

\begin{code}
G[1, \{0, 0, 1, 1, 1, 1, 1, 0, 0\}] -> \{\{1, G[1, \{1, 1, 0, 0, 1, 1, 1, 0, 0\}]\}\};

G[1, \{1, 0, 0, 1, 1, 1, 1, 0, 0\}] -> \{\{1, G[1, \{0, 1, 1, 0, 1, 1, 1, 0, 0\}]\}\};

G[1, \{1, 1, 0, 0, 0, 1, 1, 0, 0\}] -> \{\{1, G[1, \{0, 0, 1, 1, 1, 1, 0, 0, 0\}]\}\};

G[1, \{1, 0, 0, 1, 0, 1, 0, 0, 0\}] -> \{\{1, G[1, \{0, 1, 1, 0, 0, 1, 0, 0, 0\}]\}\};
\end{code}

In case one wishes to provide rules mapping an integral to multiple integrals the right-hand sides
should be presented as lists of pairs containing a coefficient and an integral. 
All rules in the file should be separated by two new lines.
This format can be also created by {\tt FIRE} automatically by the 
{\tt SaveRulesToFile[rules,filename]}, where {\tt rules} should be in a {\tt Mathematica} rules format.

Moreover, the {\tt FindRules} command can find equivalents between integrals of different families of Feynman integrals.
To do this one has to set the {\tt Problems} variable to a list of different diagram numbers,
then set the {\tt Internal[i], External[i], Propagators[i]} and {\tt Replacements[i]} for each {\tt i} from this set 
and call the {\tt FindRules} command on a set of integrals under consideration. In this case
{\tt FIRE} tries to maps integrals to the ones having higher problem numbers when possible.

The resulting file contains rules and can be loaded in {\tt Mathematica} with {\tt LoadRules[filename]} 
or in {\tt C++} with the {\tt $\#$rules} line.

\section{Internals of {\tt FIRE}}

In order to be able to set options of the {\tt C++ FIRE} properly, one should understand to some extent how it works internally.
{\tt FIRE6} launches several processes. The first among them is the {\tt FIRE6} binary itself. It reads the config file
and the files listed there, enumerates the sectors and starts running individual sector jobs. Each sector has its own database,
and unlike in older versions, there is no shared database for different sectors, so this leads to a better parallelization.
Since {\tt FIRE} needs to launch other binaries in its folder, it needs to know their location. So it should be run 
with a relative or full path, but copying the binary to other location or creating simlinks might lead to crashes.

\subsection{Reduction processes}

The {\tt FIRE6} binary first reads the list of requested integrals and puts them into corresponding sector databases. 
Then it launches individual sectors jobs ({\tt FLAME6} processes) starting from highest sectors and going down.
Multiple sector jobs can be launched at the same time in case they share the same level (number of positive indices in a sector)
and sublevel (in case of {\tt LiteRed} external symmetries the sectors which are mapped to another sector are considered to be
in a higher sublevel). The maximal number of the simultaneous sector jobs is controlled by the {\tt $\#$threads} option.

When {\tt FLAME6} works in a sector, it masks all integrals belonging to lower sectors to prevent growth of substitutions. 
They will be substituted later.

When a level (or sublevel) is over, the {\tt FIRE6} process reads the databases which were modified by the sector jobs and 
obtains the lists of integrals required in lower sectors. In also opens the lower databases and puts the required lists there, 
then proceeds with lower levels.

After {\tt FIRE} finishes with the lowest possible level, it turns back to substitutions and starts increasing levels 
launching substitution processes (also {\tt FLAME6}). In case the {\tt $\#$sthreads} option is missing, the {\tt $\#$threads}
option controls the maximum number of simultaneous jobs. Between different levels it copies required results from lower 
databases to higher databases, which does not take much computing time.

One should keep in mind that the memory usage of {\tt FIRE} is almost linearly proportional to the number of sectors 
reduced (or substituted) at the same time, so the options should be set accordingly.

There is also a mode of {\tt FIRE} that does not perform backward substitutions and creates tables listing irreducible integrals
after the forward stage. This is turned on by using {\tt $\#$masters} instead of {\tt $\#$output}.

The numbering of the sectors starts from $2$ since $1$ is reserved for the points that are of highest priority during the reduction,
meaning the right-hand of rules or integrals listed by the {\tt $\#$rules} setting. Whenever an integral 
from the right-hand side is read, {\tt FIRE} creates a relation that maps it to its virtual analogue in sector 1. 
Among other things it means that integrals is right-hands side of rules should not appear in left-hand sides for consistency. 

The {\tt $\#$preferred} setting as well as {\tt $\#$rules} setting influence the integral priority inside a sector. All integrals
listed as preferred as well as all integrals appearing in both sides of the rules have a priority in theis sectors ---
{\tt FIRE} tries to express other integrals through them if possible. 
Note: unlike in {\tt FIRE5} the corner integral in each sector is not automatically considered to be prefered. 
This might be usefull in some cases but will require adding it to prefered list if this behavior was assumed.

The sectors symmetric to lower sectors by the global symmetries do not have their own numbers,
instead the requested integrals are mapped to corresponding lowest points in orbits before reduction starts.

\subsection{Database usage and multiple computers}

{\tt FIRE6} has two ways to work with databases, the disk mode (default) and the memory mode (turned on by the {\tt $\#$memory} setting).
In the default mode {\tt FIRE} uses disk databases which are open by sector jobs or the master job for data transfer.
In the other case on-memory databases are used, and their snapshots are saved to disk when work in the sector is done.
In both cases the databases are stored in the folder specified by the {\tt $\#$database} setting.
By default it is equal to {\tt temp/db}, so one should consider whether this path is suitable for large files. 
Moreover multiple jobs using the same data path might conflict with each other.

In order to be able to recover from system crashes or timeouts, one should keep the databases. However when {\tt FIRE} starts 
working in a sector, the sector database might end in a non-consistent stage after a crash. Moreover, often one might
use temporary locations for the databases. Therefore we introduce a concept of a storage folder, another folder
where {\tt FIRE} stores copies of sector databases and copies the required databases to the database folder when needed,
then copies them back when work is done. If a storage folder is set (by the {\tt $\#$storage} setting), 
one can restart the job after a crash, and it can recover relatively fast.

The storage folder also gives the possibility to get {\tt FIRE} running on multiple nodes for the same reduction.
To do this one should start a {\tt FLAME6} binary on the slave machine with the same options.
The process will parse the configuration and then start waiting for an {\tt IP} file to appear in the storage folder.
Then the {\tt FIRE6} job starts, it opens a socket, writes its IP address into the {\tt IP} file and one of the threads
start accepting connections. The {\tt FLAME6} process from another machine connects to the open socket, 
and now {\tt FIRE6} is capable of distributing jobs both locally and remote.
For this to work the storage folder should be on a network drive with access from both nodes,
however the access to this folder is infrequent. Also the nodes should be able to connect to each other via the
TCP/IP protocol. The default port is $8080$, but can be set by the {\tt $\#$port} setting. If the port
is set to zero, this possibility is turned off.

\subsection{Parallelization}

There is a number of ways to get {\tt FIRE} to use parallelization. Some of them have been mentioned above
and are related to multiple sector jobs, both on one or multiple machines. But those are not the only options.

{\tt FIRE6} uses {\tt fermat} for calculations and sends expressions to it to get them simplified.
A number of {\tt fermat} processes are launched, defined by the {\tt $\#$fthreads} setting
(by default it is equal to {\tt $\#$threads}). There are two different ways how {\tt FIRE} can handle {\tt fermat} jobs.

The default mode is a shared pool of {\tt fermat} workers for all sectors. The child {\tt FLAME6} processes communicate
with {\tt FIRE6} via pipes and send expressions to be simplified. {\tt FIRE6} has a pool of {\tt fermat} workers
and distributes received expressions to this pool. An alternative way is to have a separate pool of {\tt fermat} workers for each 
sector job. This mode is turned on by adding the letter {\tt s} before the number in the {\tt fthreads} setting.
Please note that in this case the {\tt fthreads} is divided by the {\tt threads} setting in order
to determine the number of {\tt fermat} workers launched by a sector job. 
It depends on the problem which mode works more efficiently.

Another resource for parallelization is the usage of level threads. To use them one first needs to compile
{\tt FIRE} and dependent libraries with support of level threads ($\tt enable\_lthreads$).
Then the {\tt lthreads} setting should be used. Level threads are threads that can run in parallel in 
a given sector working almost independently for integration by parts relations of similar complexity.
By complexity of an integral inside a sector we mean a pair of numbers, the sum of positive indices minus
the number of positive indices (= number of ``dots'') and minus the number of non-positive indices
(= total power of irreducible numerators). Integrals with equal sum of numbers in this pair
can be treated almost independently, and here level threads can be applied. 
Again, the possibility to gain speedup from this feature greatly depends on the nature
of the problem and on the structure of the computers in use.
The level threads work only when the separate fermat mode is set.

Please note that the {\tt lthreads} mode (even if configured and not turned on) is slightly slower due to database locks.
This change is negligible in the polynomial version but can be noticed in the prime version, so there might be a reason
not to configure {\tt lthreads} support.

\section{Modular arithmetic}

The approach with the use of modular arithmetic\footnote{
Modular arithmetic is in use not only when solving IBP relations but also
in other problems in high-energy particle physics and quantum field theory --
see, e.g., \cite{Peraro:2016wsq}} for IBP reduction was first suggested in~\cite{vonManteuffel:2016xki}.
Basically there are two ideas here. The first one follows from the fact that we know that there are too many integration by parts relations
but before solving them we do not know any minimally required subset. So what if one first runs the reduction with
substituted variables, gets the information of what relations were really needed and then runs the real reduction?
Of course, one has to choose values for $d$ and other variables such that they do not fall into possible zeros of denominator factors.

The other idea follows from the fact that intermediate coefficients when solving sets of linear equations can be huge compared to resulting
coefficients. So one would wish to avoid this intermediate stage since it might fail to fit RAM limits.
Hence the approach is to perform reduction with different values of variables and then to reconstruct the coefficients.
However simply giving integer or rational values to all variables is normally not enough --- instead of polynomial growth 
one gets number growth. To avoid this one also sets a large prime number $p$ and used the so-called modular arithmetic
meaning all integers are replaced with their remainders from division by $p$. The set of those remainders is a field
meaning every non-zero remainder has a unique inverse. Hence moving to such a field leads to a limit for all numbers.
A prime has to be chosen big enough so it makes sense to take biggest prime numbers that fit into machine-sized 
integers ($2^{64}$) for faster calculations.

To run {\tt FIRE} with modular arithmetic one has to do the following:
\begin{itemize}
 \item Use the {\tt FIRE6p} instead of the {\tt FIRE6} binary;
 \item Set values for all variables with the {\tt $\#$variables} setting (the syntax is explained in the Appendix);
 \item Set a prime number with the {\tt $\#$prime} setting (it is not the prime number itself that is provided but its index in the set of hard-coded primes close to $2^{64}$).
\end{itemize}

A single run with the modular arithmetic might be used together with the {\tt $\#$hint} option pointing at a folder
where the hint files are to be saved. Those files list the integration by parts relations that were really used
during the reduction. A consequent run with the {\tt $\#$hint} option will get use of those files.

Now if one is going to use the modular arithmetic and reconstruction of the results, 
a reasonable number of {\tt FIRE} runs is to be made. To avoid using multiple config files,
one can use the following syntax to run {\tt FIRE6p}:

\begin{code} 
bin/FIRE6p d-p examples/doublebox
\end{code}

\noindent in the case when $d$ is the only variable and

\begin{code} 
bin/FIRE6p d-x-p examples/doublebox 
\end{code}

\noindent in the case of one extra variable. Here {\tt p} is the number of the large prime number to be used,
{\tt d} is the value of $d$ and {\tt x} is the value of the other variable.
With such a syntax one can avoid editing config files and changing the {\tt $\#$variables} and {\tt $\#$prime} settings.

More variable substitutions can also be used although there is no stable reconstruction procedure at this point. The usage is the following:
\begin{itemize}
 \item Variable values should be separated by -;
 \item No negative values allowed;
 \item Last number indicates the prime number;
 \item Variables are replaced in the order they appear in the {\tt $\#$variables} section, first $N$ variables are replaced where $N$ is the number of values provided in the command line minus one;
 \item Replacements like {\tt d->57} in the configuration file get overwritten if they are among the first $N$ variables;
 \item Replacements like {\tt d->/57} in the configuration file get joined and form a fraction;
 \item Extra variables after the first $N$ should have replacement values in the {\tt $\#$variables} section.
\end{itemize}

One might need a huge number of jobs like that, and consequent runs can take too much time. Therefore this approach is
most suitable when a supercomputer is used. To use massive parallelization one should prepare a config file and then
use the {\tt FIRE6\_MPI} binary. It has the following syntax:

\begin{code} 
bin/FIRE6\_MPI -c config p d\_min d\_max 
\end{code}

\noindent in the case when $d$ is the only variable and

\begin{code} 
bin/FIRE6\_MPI -c config p d\_min d\_max x\_min x\_max 
\end{code}

\noindent in the case of one extra variable. Here {\tt p} is the number of different primes that will be used (starting from the biggest
possible prime number corresponding to the setting {\tt $\#$prime 1} and going on), {\tt d\_min} is the minimal value of $d$,
{\tt d\_max} is the maximal value of $d$, similarly for {\tt x\_min} and {\tt x\_min}.
The name of the other variable in problem files is unimportant.

Currently the MPI version is uncapable of running jobs with more than two variables to be substituted with different values.

As a result {\tt p * (d\_max - d\_min + 1) * (x\_max - x\_min + 1)} jobs are to be run one after another.
However the {\tt FIRE6\_MPI} should not be called directly but through the MPI system of the supercomputer.
For instructions one should refer to the instructions of the particular cluster. As a result
one runs a large number of jobs at the same time.

It is safe to use one config file and have multiple jobs running at the same machine through MPI: {\tt FIRE} creates
subfolders in the database folder with names containing process id's and node names for databases. However in case of crashes
(which is normal for supercomputers) those folders might be remaining, so sometimes one should clean then not to run out of quota.

As a result each {\tt FIRE} job creates a tables file, but with a changed name: {\tt name-d-p.tables} or {\tt name-d-p-x.tables}.
Moreover, each job ``reserves'' tables files when starts working by creating empty tables with such names, so that one can launch
multiple MPI jobs at the same time. Again, in case of supercomputer node crashes one should delete empty tables files.
This can be done by

\begin{code}
 find FOLDERNAME -size  0 -print0 |xargs -0 rm $--$
\end{code}

The jobs should be run until all tables are created. And then one can move to the coefficient reconstruction.

One can also run multiple tasks without MPI but for example with the use of GPU parallel, for example,

\begin{code}
parallel "bin/FIRE6p -variables \{1\}-\{2\} -c config" ::: $\$$(seq 100 110) ::: $\$$(seq 1 3)  
\end{code}

However a direct call to such a command will result in a crash because the jobs conflict with other (databases and semaphores) 
and also can provide too much output. So the command should be changed to 

\begin{code}
parallel "bin/FIRE6p -variables \{1\}-\{2\} -c config -parallel -silent" ::: $\$$(seq 100 110) ::: $\$$(seq 1 3)  
\end{code}

\noindent to enable the concurent mode with no conflic between individual jobs (process id's are added to the database paths and POSIX named semaphores are also modified).

The recommended options for running multiple modular jobs are:
\begin{itemize}
 \item {\tt $\#$memory} for faster runs;
 \item {\tt $\#$compressor none} in case there is enough RAM;
 \item {\tt $\#$bucket N} where {\tt N} is the smallest number that does not result in ``reopening database with bucket... '' messages;
 \item {\tt $\#$wrap} if there is a quota on the number of files;
 \item {\tt $\#$small} to save some RAM.
\end{itemize}

Still before running jobs one does not know how many tables will be required. We will try to give our ``algorithm'' on how
those jobs should be run, but let us first explain the reconstruction process.

\subsection{Reconstruction}

The reconstruction of coefficients consists of two stages. First there is the reconstruction of rational coefficients from modular numbers
(while variables have still fixed values). It searches for the rational number with minimal absolute values of the numerator and denominator 
that is mapped to the obtained remainders over large prime numbers. When increasing the number of primes no longer 
changes the result, some more checks are made and the rational number is reconstructed. {\tt FIRE} can perform this operation
on all coefficients in tables, this is done by the {\tt Mathematica} part of {\tt FIRE6} with:

\begin{code}
RationalReconstructTables[filename, pnum]
\end{code}
 
\noindent command. Here {\tt filename} is the name of the target filename (without the {\tt p-} part) and {\tt pnum}
is the number of different prime numbers in use, starting from the first (largest). The result is the 
requested tables file if successful. {\tt FIRE} produces messages which can be turned off with
adding the {\tt True} option:

\begin{code}
RationalReconstructTables[filename, pnum, True]
\end{code}

\subsection{One-dimensional reconstruction}

When the rational reconstruction is done, one can move to the reconstruction of coefficients. Here we recall classical 
approaches to the reconstruction of formulas. Let us start with the case of one variable. 
One of the approached is based on the so-called Newton reconstruction which uses the the following formula:

\begin{code}
a[0] + (x - x[0]) (a[1] + (x - x[1]) ( $\ldots$ a[n] )$\ldots$ ))
\end{code} 

There are well-defined algorithms that can perform this reconstruction for any polynomial (not a rational function).
If a reconstruction of a rational function is needed, then one should use the Thiele formula:

\begin{code} 
a[0] + (x - x[0]) / (a[1] + (x - x[1]) / ( $\ldots$ a[n]) $\ldots$ )) 
\end{code}

One should keep in mind that this method needs more values. For example, reconstructing a polynomial with the Newton formula needs
the number of points close to its power, but reconstructing the same polynomial with the Thiele formula
needs about twice that many points.

One can perform the reconstruction with the 

\begin{code} 
ThieleReconstructTables[filename, dlist] 
\end{code}

Here {\tt filename} is the reconstruction target filename and {\tt dlist} is the list of different values of $d$
that will be used. As a result one obtains the file with tables in the same format, but containing rational polynomial
coefficients. {\tt FIRE} produces messages which can be turned off with
adding the {\tt True} option.

Hence the general instructions to perform a one-dimensional reconstruction are the following

\begin{enumerate}
 \item Choose a relatively big number of $d$ and run modular arithmetic reductions until the rational numbers can be reconstructed (for example, in \cite{Lee:2019zop} we required $13$ values);
 \item Increase the number if primes used a bit to be safe and run massive reductions at the supercomputer with different values of prime numbers and $d$;
 \item Run the rational reconstruction for each value of $d$ in use, if it fails, increase the number of primes and return to step 2;
 \item Run the Thiele reconstruction, if it fails, increase the number of values of $d$ and return to step 2.
\end{enumerate}

\subsection{Two-dimensional reconstruction}

Now when considering the case of two variables there is nothing known to work better than a mixture of the two approaches.
One builds a Newton or Thiele formula in one variable and the coefficients are build by the
Newton or Thiele formula in another variable. Now the problem is that a Thiele-Thiele formula 
might make the reconstruction too complex, both computationally and also requiring a large number of tables,
but other variants do not cover the general case of a rational function of two variables.
However, we can take into account that the denominators in IBPs correspond
to singularities of Feynman integrals which correspond to thresholds and poles
whose positions are described in terms of kinematical invariants and masses,
so that the dependence on $d$ should not mix with the dependence on these variables.
In practice, it can happen however that these variables do mix with $d$.
Then the natural procedure is to find a basis
of master integrals such that the denominators are split into a product of a function of $d$ and a function of $x$. 
In our experience this is always possible.
We will call such a basis \textit{factorising}.
It can be checked that a basis is factorising on a simpler task that can be reduced with the classical approach.

So now supposing we have a factorising basis, we can continue in the following way. First, after the numerical reconstruction
of rational functions is complete, one fixes a value of $x$ and recovers the tables with the one-dimensional Thiele function.
Then the same is done with fixing a value of $d$. Now knowing the structure of denominators we understand that 
we can take the least common multiple of tables with $d$ and tables with $x$, multiply those factors, and 
afterwards all coefficients multiplied by this factor become polynomials both in $x$ an $d$.
Then one can run a Newton-Newton reconstruction. 

Now let us proceed to instructions on how one uses this approach provided a factorising basis has been found.

\begin{enumerate}
 \item Fix a value of $d$ and a value of $x$ (we will use this name for the variable) and build a number of tables 
 until rational reconstruction can be performed with

\begin{code}
RationalReconstructTables["TASK-" <> ToString[d0] <> \\"-" <> ToString[x0] <> ".tables", p0, True]
\end{code}

{\tt p0} stands for the number of different big primes in tables, {\tt True} stands for silent, giving only a summary for each table reconstruction.
If "Rational reconstruction unstable" is encountered, {\tt p0} has to be increased. After that one knows the number of prime numbers 
needed for the reconstruction. In the following steps this should be increased by $1$ or $2$ to be on the safe side;

\item Fix a value of x and create a number of tables with different values of $d$ and $p$ until Thiele reconstruction over $d$ succeeds with

\begin{code}
For[x = x0, x <= x1, ++x,\\
  RationalReconstructTables[
    "TASK-" <> ToString[d0] <>  "-" <> ToString[x] <> ".tables", p0, True]\\
];\\
ThieleReconstructTables["TASK-"<>ToString[d0]<> ".tables", Range[x0, x1], True]
\end{code}

Here $d0$ is the chosen value for $d$, $x0$ is the minimal $x$ and $x1$ is the maximal $x$ for which there are tables,
{\tt True} again stands for the silent mode. If the reconstruction is unstable, $x1$ should be increased.
When the Thiele reconstruction is stable, the required denominator factor can be recovered from new tables by

\begin{code} 
DenominatorFactor["TASK-"<>ToString[d0]<>.tables"] /.{d->x} 
\end{code}

The replacement here is needed because the Thiele reconstruction originally is intended for
the case of one variable and assumes it is $d$. The answer returned is the worst denominator and will work 
as the factor later. To be safe one can recheck the factor with another d value reconstruction.

Note that the Thiele reconstruction in $x$ needs more values that the later appearing Newton reconstruction
after multiplying by a coefficient. So when there is the factor, the reconstruction should be rerun with

\begin{code} 
NewtonReconstructTables[TASK <> "-" <> ToString[d0] <> \\".tables", Range[x0, x1], factorX, True];
\end{code}

This gives the information 
on how many values of $x$ for the final table creation, and it is normally much smaller than the 
original number needed for the Thiele reconstruction.

\item A similar procedure should be repeated with a fixed value of $x$ and running

\begin{code}
For[d = d0, d <= d1, ++d,\\
  RationalReconstructTables["TASK-" <> ToString[d] <> "-" <> ToString[x0] <> ".tables", p0, True]\\
];\\
ThieleReconstructTables[TASK, Range[d0, d1], True, "-" <> ToString[x0] <> ".tables"];
\end{code}

Note that there is a change in the {\tt ThieleReconstructTables} syntax. There is an extra argument,
that is appended to the tables name after the running index.

After the reconstruction goes through, the worst factor in $d$ is obtained with
{\tt DenominatorFactor["TASK.tables"]}. The factor and the number of different $x$ values can
be checked and obtained with

\begin{code}
NewtonReconstructTables[TASK <> ".tables", Range[d0, d1], factorD, True, "-" <> ToString[x0] <> ".tables"];
\end{code}

\item Now one knows the worst possible denominator which is a product of the factor in $d$ and the factor in $x$,
one also knows the number of different values of $d$ and the number of different valuer of $x$ required for the reconstruction.
This number of tables must now be created.

The final reconstruction is run with

\begin{code}
For[d = d0, d <= d1, ++d,\\
 For[x = x0, x <= x1, ++x,\\
  RationalReconstructTables[
    "TASK-"<>ToString[d]<>"-"<>ToString[x]<>".tables", p0,True]
  ]
];\\
NewtonNewtonReconstructTables["TASK.tables", Range[d0, d1], Range[x0, x1], x, denominator, False];
\end{code}

As a result the tables with reconstructed coefficients are created.

\end{enumerate}

\section*{Appendix: config files for C++}

Most of {\tt FIRE} options are provided via the config files. But a few can come via the command line:

\begin{itemize}
 \item {\tt -c filename} --- the path to the config file with no extension
 \item {\tt -variables vars} (only prime mode) --- lists the substitution values for variables separated by ``{\tt -}'', the last one is the index of the prime number to be used;
 \item {\tt -parallel} --- makes {\tt FIRE} use modified paths and named semaphores, so that multiple jobs do not conflict with each other;
 \item {\tt -silent} --- turns off most of the output;
 \item {\tt -database path} -- same with the {\tt $\#$database} option in the config file; usefull if a dynamic path is to be specified; this option overrided the {\tt $\#$database} entry in the config file if both are present;
 \item {\tt -sector s} (option of {\tt FLAME}) --- a direct way to make the {\tt FLAME6} work in a particular sector; positive values mean forward pass, negative values mean backward pass;
\end{itemize}

Let us now summarize all options of config files. The order of some of the options is important,
so it is recommended to keep the order as described here or in supplementary examples (especially the order of the options that are listed after the {\tt $\#$start} command). A line starting with $\#\#$ is considered as a comment and is ignored by {\tt FIRE}.

\begin{itemize}

 \item {\tt $\#$fermat} (optional) --- the path to the {\tt fermat} binary. By default the binary shipped with {\tt FIRE} is used, but one might wish to change it;
 \item {\tt $\#$compressor} (optional) --- compressor choice for the database engine. The possible compressors depend on compilation options. The full set of values is {\tt lz4, lz4fast, lz4hc, zlib, snappy, zstd, none}. The first three are from the {\tt lz4} family with {\tt lz4hc} being the slowest and compressing most and {\tt lz4fast} being the fastest among all compressors shipped with {\tt FIRE}. {\tt lz4fast} is the suggested option for the prime version, probably together with the {\tt $\#$small} option. For a long time we used {\tt snappy} as the best choice for the polynomial version as being able to compress well enough but reasonably fast. However there is a recent addition to {\tt FIRE}, the {\tt zstd} option standing for the ZStandard compressor that claims to compress better than snappy at the same speed. The {\tt none} setting means no compressor and can provide a noticable speedup for the prime version in case there is enough RAM;
 \item {\tt $\#$threads} --- the number of threads launched for parallel reduction of sectors of same level; 
 \item {\tt $\#$fthreads} (optional) --- the number of {\tt fermat} processes launched. By default it is equal to the number of threads, but it might make sense to increase it; an {\tt s} before the number turns on the separate fermat mode;
 \item {\tt $\#$sthreads} (optional) --- the number of threads used during the substitution stage. By default it is equal to the number of threads, but it might make sense to decrease it in case of problems with RAM;
 \item {\tt $\#$lthreads} (optional) --- the number of level threads ({\tt FIRE} should be compiled with support of level threads for this option);
 \item {\tt $\#$port} (optional) --- the port for the main job to listen to accept child connections. $0$ is the default value meaning the setting if off, value $8080$ is recommended to enable;
 \item {\tt $\#$variables} --- comma-separated list of variables used during the reduction; also this setting can provide variable replacements like {\tt d->57}, the right-hand sides of rules should be numbers;
 \item {\tt $\#$pos\_pref} (optional) --- a setting that allows to tweak the choice of master integrals. By default {\tt FIRE} chooses such an ordering that having one positive shift has a higher priority that having a negative shift, meaning that obtains master integrals with one $2$ instead of one $-1$. The integrals with positive indices are easier in a sense because the equivalents between them can be found. But in cases where there are many masses sometimes this priority is not enough, and one gets integrals with two or more $-1$. To set a priority for integrals with two (or more) ``dots'', one has to increase the {\tt pos\_pref} setting (by default $1$). This also leads to internal symmetries being applied to a larger subset of integrals; 
 \item {\tt $\#$database} (optional) --- path to the place where {\tt FIRE} stores the data; by default it points to the {\tt temp/db} directory relative to the current folder, but this can be changed; it is especially important to be sure that this setting does not point to a network drive in case one does not use the {\tt memory} setting;
 \item {\tt $\#$storage} (optional) --- sets a folder for the storage of database files; by default, the storage is used only at the forward pass, to use it also during substitutions preceede the path with an exclamation mark (``!'', no quotes); please note that this option is incompatible with different {\tt $\#$threads} and {\tt $\#$sthreads} -- {\tt FIRE} cannot distnguish properly local and remote threads when some are to be stopped;
 \item {\tt $\#$bucket} (database tuning) --- an integer number equal to $20$ by default and related to the database engine; small values make {\tt FIRE} auto-increase the bucket during the reduction, and this can slow things down, large values can make {\tt FIRE} use too much RAM; for complicated tasks consider a bucket value equal to $27$--$30$; the old values are remaining here for backward compatibility, but currently {\tt FIRE} uses the shifted (by $-4$) value and prints it in the log;
 \item {\tt $\#$wrap} (database tuning) --- results in sector databases being stored in a general database, only the active databases remain being files; this greatly reduces the number of files used which can be important in case of file limits when running multiple jobs on a cluster;
 \item {\tt $\#$prime} (prime version) --- is used and should be used only in the ``prime'' version of {\tt FIRE}. It sets the prime number to be used in evaluations. If the option is set to zero, the prime number $2017$ is used which is useful for tests (to see small numbers in results). Any other value uses one of the maximal prime numbers fitting into a 64-but unsigned integer (currently there are $128$ hard-coded primes in {\tt FIRE});
 \item {\tt $\#$memory} (optional) --- if this line exists, {\tt FIRE} uses the ``RAM mode'' instead of the ``disk mode'' --- it stores active databases in RAM; this setting makes {\tt FIRE} use more RAM, but it becomes less vulnerably to freezing because of network drives;
 \item {\tt $\#$clean} (optional) --- cleans the POSIX named semaphores opened by older runs of {\tt FIRE} that crashed; safe unless one user launches multiple {\tt FIRE} jobs at the same machine simultaneously; this might be required when {\tt FIRE} starts crashing at random places at a particular machine that there is no way to reboot;
 \item {\tt $\#$keepall} (debugging) --- keeps all entries in all databases, no removal;
 \item {\tt $\#$small} (database tuning) --- decreases the key size for database entries, which is especially efficient in case of the number of indices about $10$. This might make sense to decrease database size in the ``prime'' mode;
 \item {\tt $\#$allIBP} (for non-standard IBPs) --- by default {\tt FIRE} does not use some of the integration by parts relations that are guaranteed consequences of others; this logic is valid in case those are real IBPs in the classical case but might result in missing some relations in special cases. If this is the situation, then this line has to be a part of the config file;
 \item {\tt $\#$nolock} (database tuning) --- if this line exists, {\tt FIRE} does not lock databases in ``disk mode''; this might be required to get {\tt FIRE} work on some file systems;
 \item {\tt $\#$start} --- just a command following the previous lines;
 \item {\tt $\#$folder} (optional) --- if this path is given, all following paths will be considered relative to this folder unless they are absolute paths (starting with {\tt $/$});
 \item {\tt $\#$problem} --- the instruction to load a start or sbases file; the syntax is {\tt $\#$problem pn filename} or {\tt $\#$problem pn |maxpos|filename} or {\tt $\#$problem pn |minpos,maxpos|filename}; {\tt pn} here is the diagram number; if {\tt maxpos} is provided, then indices bigger than {\tt maxpos} cannot be positive; if {\tt minpos} is provided, then indices smaller than {\tt minpos} cannot be positive;
 \item {\tt $\#$hint} (optional) --- points to a folder with hint files which list the relations that should be used (or created if missing); might be useful if one first makes a ``prime'' version run, then uses the obtained hint files; 
 \item {\tt $\#$lbases} (optional) --- a command to load with {\tt LiteRed} rules obtained by {\tt TransformRules};
 \item {\tt $\#$output} or {\tt $\#$masters} --- the path where {\tt FIRE} will store the resulting tables; if one chooses the {\tt $\#$masters} syntax, then {\tt FIRE} only aims at finding master integrals, and this can be much faster than the whole reduction; this might be needed to find master integrals, then one can use {\tt WriteRules} to find equivalents between them, so that afterwards the full reduction can be run with the use of those rules;
 \item {\tt $\#$preferred} (optional) --- this file can list integrals that are preferred as master integrals; this might be needed if one does not like the automatic choice; however, one should keep in mind that it is just a hint for {\tt FIRE}; the syntax of the file is the same with the input list;
 \item {\tt $\#$rules} (optional) --- a command to load a file with rules for some integrals; the syntax is explained in section~\ref{rules};
 \item {\tt $\#$integrals} --- the file with integrals to be reduced.

\end{itemize}

\section*{Conclusion}

We presented a new version of the {\tt FIRE} program performing Feynman integral reduction. The new version has been developed in private for a few years and was successfully used in multiple projects. Currently it is made public of the bitbucket. The new version offers a lot of functionality requested by the users and by the needs of reduction problems --- the possibility to recover from crashes, usage of multiple machines for one reduction, parallelization inside sectors, modular arithmetic, support of supercomputers and other features. We hope that the new public version will be widely used in modern research in elementary particle physics. We have more plans on how {\tt FIRE} can be improved and we are ready to accept improvement ideas and error reports.

\section*{Acknowledgements}

The work is partially supported by RFBR, grant 17-02-00175. The authors would like to thank 
Vladimir Smirnov, Mathias Steinhauser, Joshua Davies and Go Mishima
for useful advices, discussions and testing the program in its development stage.
Testing of the MPI version of {\tt FIRE} was performed at the MSU Lomonosov supercomputer\cite{opanasenko2013lomonosov3827487}.

\bibliographystyle{elsarticle-num-names}
\bibliography{FIRE-new,asmirnov}
\end{document}